\documentclass[conference, 10pt]{IEEEtran}
\IEEEoverridecommandlockouts
\usepackage{cite}
\usepackage{amsmath,amssymb,amsfonts}
\usepackage{algorithmic}
\usepackage{graphicx}
\usepackage{textcomp}
\def\BibTeX{{\rm B\kern-.05em{\sc i\kern-.025em b}\kern-.08em
    T\kern-.1667em\lower.7ex\hbox{E}\kern-.125emX}}
\usepackage{algorithm}
\usepackage{array}
\usepackage{stfloats}
\usepackage{url}
\usepackage{verbatim}
\usepackage{amsmath}
\usepackage{subcaption}

\usepackage{hyperref}

\newcommand{\refappendix}[1]{\hyperref[#1]{Appendix~\ref*{#1}}}
\usepackage{booktabs}
\DeclareMathOperator{\sign}{sign}

\usepackage[table]{xcolor}
\usepackage{soul}

\definecolor{plot0}{HTML}{004488}
\definecolor{plot1}{HTML}{DDAA33}
\definecolor{plot2}{HTML}{BB5566}
\definecolor{plot3}{HTML}{000000}
\definecolor{plot4}{HTML}{AAAAAA}

\usepackage{siunitx}
\usepackage{pgfplots}
\pgfplotsset{compat=1.18}
\usepackage{balance}
\usetikzlibrary{shapes.geometric}

\usepackage{pgfkeys}
\usepackage{float}
\usepackage{multirow}
\usepackage{collcell}

\pgfkeys{/heat/.is family, /heat,
    min colour/.initial = white,
    max colour/.initial = plot0,
    text colour/.initial = black,
    min color/.style = {min colour=#1},
    max color/.style = {max colour=#1},
    text color/.style = {text colour=#1},
    min/.initial = 0,
    max/.initial = 50,
    slider/.code={%
       \tikz{\shade[left color=\HVal{min colour},%
                    right color=\HVal{max colour}]%
          (current page.south west) rectangle ++(#1,12pt);
       }%
    }%
}

\newcommand\HVal[1]{\pgfkeysvalueof{/heat/#1}}

\newcolumntype{H}{>{\collectcell\Heat}r<{\endcollectcell}}
\newcommand\Heat[1]{
  \pgfmathparse{int(100*(#1-\HVal{min})/(\HVal{max}-\HVal{min}))}
  \edef\HeatCell{\noexpand\cellcolor{\HVal{max colour}!\pgfmathresult!\HVal{min colour}}}%
  \HeatCell\textcolor{\HVal{text colour}}{$#1$}%
}

\newcommand{\INDSTATE}[1][1]{\STATE\hspace{#1\algorithmicindent}}

\begin{document}

\title{Forward-Forward Autoencoder Architectures for\\Energy-Efficient Wireless Communications
\thanks{This work was supported in part by the German Research Foundation (DFG) as part of Germany’s Excellence Strategy - EXC 2050/2 - Project ID 390696704 - Cluster of Excellence \emph{Centre for Tactile Internet with Human-in-the-Loop} (CeTI) of Technische Universität Dresden, by the German Federal Ministry of Research, Technology and Space (BMFTR) within the national initiative on 6G Communication Systems through the transfer hubs \emph{6G-life} under Grant 16KIS2413K and \emph{6GEM+} under Grants 16KIS2412 and 16KISS005, and by the ZENITH Research and Leadership Career Development Fund, Swedish Foundation for Strategic Research (SSF). This work is based on research activities within the COST Action 6G-PHYSEC (CA22168).
}}

\author{\IEEEauthorblockN{Daniel Seifert\textsuperscript{$*$}, Onur Günlü\textsuperscript{$\dagger$$\ddagger$}, and Rafael F. Schaefer\textsuperscript{$*$}\\[1ex]}
\IEEEauthorblockA{\textsuperscript{$*$}Chair of Information Theory and Machine Learning, Technische Universität Dresden, Germany \\
\textsuperscript{$\dagger$}Lehrstuhl für Nachrichtentechnik, Technische Universität Dortmund, Germany \\
\textsuperscript{$\ddagger$}Information Theory and Security Laboratory (ITSL), Linköping University, Sweden\\[0.5ex]
\{daniel.seifert, rafael.schaefer\}@tu-dresden.de, onur.guenlue@tu-dortmund.de}
\vspace{-.9cm}
}

\maketitle

\begin{abstract}
The application of deep learning to the area of communications systems has been a growing field of interest in recent years. Forward-forward (FF) learning is an efficient alternative to the backpropagation (BP) algorithm, which is typically used as the training procedure for neural networks. Among its several advantages, FF learning does not require the communication channel to be differentiable and does not rely on the global availability of partial derivatives, allowing for an energy-efficient implementation. In this work, we design end-to-end learned autoencoders using the FF algorithm and numerically evaluate their performance for the additive white Gaussian noise and Rayleigh block fading channels. We demonstrate their competitiveness with BP-trained systems in the case of joint coding and modulation, and in a scenario where a fixed, non-differentiable modulation stage is applied. Moreover, we provide further insights into the design principles of the FF network, its training convergence behavior, and significant memory and processing time savings compared to BP-based approaches.
\end{abstract}


\vspace{-0.15cm}
\section{Introduction}
The backpropagation (BP) algorithm~\cite{rumelhart1986} is the main enabler of the tremendous success of the application of neural networks to problems across various research fields in recent years. Thus, it is the default algorithm for optimizing the network parameters during training. In the field of communications, deep learning-based approaches aim to overcome the suboptimality originating from inadequate mathematical modeling and block-wise processing~\cite{oshea2017, dornerDeepLearningBased2018}. Moreover, they are envisioned to improve reliability by rapidly adjusting to changing environmental conditions that affect the link quality~\cite{aoudia2019}. However, BP has certain properties that make its deployment in communications systems difficult. 

Firstly, the algorithm requires a fully differentiable path through the neural network. For instance, when deploying end-to-end learned coding schemes, this prerequisite can be fulfilled only in theory by resorting to simplified channel models, whereas in real-world channels, we can only observe input and output samples. In~\cite{aoudia2019}, a framework has been proposed that relies on reinforcement learning (RL) to train the transmitter and receiver separately. The estimation of the gradients in the transmitter is enabled by an additional noiseless feedback link, over which the receiver's loss is shared. Moreover, the channel can be modeled by generative approaches, such as generative adversarial networks (GANs)~\cite{dornerWGANbasedAutoencoderTraining2020} or diffusion models~\cite{kimRobustGenerationChannel2024a}, which are differentiable by definition.

Another drawback of BP is its inefficiency in terms of memory and energy. The state-of-the-art hardware implementations of neural networks typically involve graphics processing units (GPUs) and application-specific integrated circuits (ASICs), which operate in the digital domain. Recent shifts towards neuromorphic and fully analog hardware are challenged by BP's memory consumption due to the necessity of storing the partial derivatives of each function or node in the backward pass. Ex-situ and hybrid approaches typically train the neural network externally and then map the resulting weights onto analog hardware, such as memristors~\cite{aguirreHardwareImplementationMemristorbased2024}. Due to hardware imperfections, the performance can significantly differ from the software implementation. The desirable in-situ training could be achieved efficiently by calculating the weight changes layer-wise based on the forward and the backward (error) signal~\cite{vandoremaele2024}. However, this technique would again rely on the availability of a fully differentiable backward path.

Finally, the BP algorithm leads to several forms of locking mechanisms, the most crucial lying in the backward path, i.e., all layers have to wait until the gradient calculation of their corresponding successor has finished. \cite{jaderbergDecoupledNeuralInterfaces2017a} proposes a framework for decoupling subsets of neural networks using synthetic gradient models to overcome this problem. However, this approach carries a processing and memory overhead, as the approximated gradients still need to be tracked and stored.

To address some of these challenges, various forward-only algorithms have been proposed, mainly motivated by the biological implausibility of BP. However, only a few achieve the same accuracy as BP~\cite{journeHebbianDeepLearning2023}. In contrast, more advanced Hebbian learning approaches based on different neural plasticity rules are able to narrow this gap~\cite{journeHebbianDeepLearning2023}. In particular, the change of weights between two neural layers is determined by common excitation and, thus, does not require any feedback signal. Another learning framework, called~\emph{sigprop}~\cite{kohanSignalPropagationFramework2024}, is based on the propagation of a learning signal in parallel to the data path and requires a separate representation of data and label in all hidden dimensions, allowing for layer-wise training.

The forward-forward (FF) algorithm~\cite{hinton2022} focuses on a similar yet more general idea that performs two forward passes on the neural network and adjusts its parameters with respect to a goodness metric in each layer. Although FF showed inferior performance in comparison to BP during initial experiments on classification tasks using the MNIST and CIFAR-10 datasets~\cite{hinton2022}, it possesses highly appealing properties for applications in communications systems: It does not require a differentiable channel as gradients are not backpropagated through the whole network. In addition, the algorithm could be implemented in fully analog, low-power circuits, as it is composed of simple operations. Since there is no need to store the derivatives in each layer, it would also operate more efficiently with respect to its memory usage. Moreover, it allows for a pipelined training procedure, since layer-wise parameter tuning eliminates the backward lock. These beneficial algorithmic properties have recently facilitated the first memristor-based hardware implementation of in-situ FF learning, using $460$ times less energy than comparable ex-situ trained memristive approaches~\cite{renaudineauForwardonlyLearningMemristor2026}.

In this work, we propose a design for end-to-end learned autoencoders for communications that are trained with the FF algorithm using contrastive input data. We numerically evaluate the performance of these systems in terms of the block error rate (BLER) and draw comparisons with BP-based approaches for the scenarios of joint coding and modulation, as well as for the case of enforcing a modulation on the encoder output. In addition, we quantify the relation between network size and BLER for FF autoencoders. We observe that the FF autoencoder can achieve performance close to that of BP and even surpass it when non-differentiable operations are involved. Furthermore, we examine the convergence rate of the proposed design and, considering comparable-performance networks, show that FF autoencoders are also able to compete with and even outperform their BP-based counterparts. Finally, we highlight processing time and memory savings that originate from the algorithmic properties of FF learning.

\section{Forward-Forward Autoencoder}
\begin{figure*}
    \centering
    \includegraphics[width=0.92\textwidth]{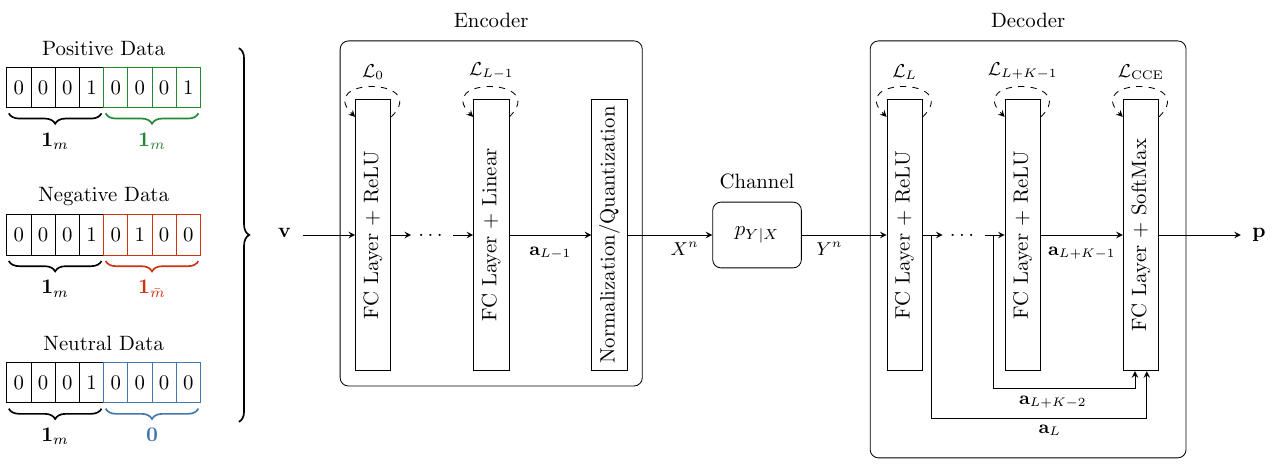}
    \caption{Autoencoder architecture trained with the FF algorithm, where each layer can employ an individual loss function.}
    \label{fig:ff_autoencoder}
    \vspace{-.5cm}
\end{figure*}
The autoencoder is a neural network composed of an encoder-decoder pair that is jointly optimized to find efficiently coded representations of the input data and retrieve the original data from the coded examples. Specifically, the configuration of overcomplete autoencoders, i.e., the encoded bottleneck has a greater dimensionality than the input data, shows a resemblance to the problem of channel coding in communications, where redundancies are actively introduced to make the transmission more robust against perturbations introduced by the channel~\cite{oshea2017}. Given the message~$m\in\mathcal{M}=\{0,\ldots,2^k-1\}$ whose binary representation consists of~$k$ bits, the input to the autoencoder is typically transformed into a one-hot representation~$\mathbf{1}_m\in\{0,1\}^q$, i.e., a zero vector of length~$q=2^k$ with a single one at index~$m$. At the output of the network, a softmax function estimates the probability vector~$\mathbf{p}\in\left[0,1\right]^q$ for the decoded message. Consequently, the categorical cross-entropy (CCE) loss is applied to quantify the difference between the two vectors. This loss function inherently optimizes the autoencoder with respect to the BLER~\cite{wiesmayrBitErrorBlock2023}.

In this work, we consider networks that consist of simple multilayer perceptrons (MLPs), i.e., fully-connected (FC) layers as displayed in Fig.~\ref{fig:ff_autoencoder}. These layers are characterized by a set of learnable parameters~$\boldsymbol{\theta}_i$ and activation functions which are, except for the last encoder layer, implemented by non-linear rectified linear units (ReLU). The corresponding output of each layer~$i$ is denoted by~$\mathbf{a}_i=f_{\boldsymbol{\theta}_i}(\mathbf{x}_i)$ where~$\mathbf{x}_i$ is the input to the layer. Moreover, the encoder deploys a normalization or quantization stage to ensure an average or hard power constraint, respectively, before the encoded block is transmitted over the channel~$p_{Y|X}$.

In contrast to the BP algorithm, FF learning adjusts the parameters of the neural network in two forward passes --- one with positive, another with negative data. The FF algorithm does not perform a backward pass through the whole network, propagating the derivatives originating from the loss function, which compares the network output with the ground truth. Thus, the labels must be included in the input to the network. In contrast to previous work using the MNIST dataset~\cite{hinton2022} where the labels were encoded in the first pixels of the input image to the network, for our autoencoder design we propose to assemble the input to the network~$\mathbf{v}$ in a contrastive manner: Positive samples are generated by simple replication, i.e., $\mathbf{v}=(\mathbf{1}_m|| \mathbf{1}_m)$ where~$(\cdot || \cdot)$ denotes concatenation. A negative sample is composed by randomly sampling a second message~$\bar{m}\in\mathcal{M}$ with $\bar{m}\neq m$ and concatenating its one-hot representation to the true message vector such that~$\mathbf{v}=(\mathbf{1}_m|| \mathbf{1}_{\bar{m}})$. For inference, we use a neutral label such that the input vector to the encoder is given by~$\mathbf{v}=(\mathbf{1}_m|| \mathbf{0})$, where~$\mathbf{0}$ denotes the all-zeros vector. Note that due to these introduced zeros, the corresponding input nodes would effectively be deactivated.

\begin{figure}[t]
\vspace{-0.2cm}
\begin{algorithm}[H]
\caption{FF Autoencoder Training}\label{alg:training}
\begin{algorithmic} \small
\INDSTATE[0]\textbf{Instantiate}: encoder~$\boldsymbol{\theta}^\mathrm{e}$, decoder~$\boldsymbol{\theta}^\mathrm{d}$, classifier~$\boldsymbol{\kappa}$
\INDSTATE[0]\textbf{repeat}
\INDSTATE[1]$\text{sample random } m \in \mathcal{M} $
\INDSTATE[1]$\triangleright$ FF networks: positive pass
\INDSTATE[1]$(\mathcal{L}_0^+,\dots,\mathcal{L}_{L+K-1}^+,\mathcal{L}_\mathrm{CCE}^+) \gets$ \textsc{FF-AE}($m$, positive)
\INDSTATE[1]$\triangleright$ FF networks: negative pass
\INDSTATE[1]$(\mathcal{L}_0^-,\dots,\mathcal{L}_{L+K-1}^-,\mathcal{L}_\mathrm{CCE}^-) \gets$ \textsc{FF-AE}($m$, negative)
\INDSTATE[1]$\triangleright$ Classifier: neutral pass
\INDSTATE[1]$(\mathcal{L}_0,\dots,\mathcal{L}_{L+K-1},\mathcal{L}_\mathrm{CCE}) \gets$ \textsc{FF-AE}($m$, neutral)
\INDSTATE[1]$\triangleright$ Optimizer step for FF networks:
\INDSTATE[1]$\mathrm{SGD}((\boldsymbol{\theta}^\mathrm{e}, \boldsymbol{\theta}^\mathrm{d}), (\mathcal{L}_0^+,\dots,\mathcal{L}_{L+K-1}^+,\mathcal{L}_0^-,\dots,\mathcal{L}_{L+K-1}^-),$
\INDSTATE[2]$\qquad\gamma_f, \lambda_f, \mu_f)$
\INDSTATE[1]$\triangleright$ Optimizer step for classifier:
\INDSTATE[1]$\mathrm{SGD}(\boldsymbol{\kappa}, \mathcal{L}_\mathrm{CCE}, \gamma_c, \lambda_c, \mu_c)$
\INDSTATE[0]\textbf{until} stop criterion is met
\end{algorithmic}
\label{alg1}
\end{algorithm}
\vspace{-0.4cm}
\begin{algorithm}[H]
\caption{FF Autoencoder}\label{alg:autoencoder}
\begin{algorithmic} \small
\INDSTATE[0]\textbf{function}~\textsc{FF-AE}($m$, $t$):
\INDSTATE[1]\textbf{switch} $t$ \textbf{do}
\INDSTATE[2]\textbf{case} positive:~$\mathbf{v} \gets \left( \mathbf{1}_m || \mathbf{1}_m\right)$
\INDSTATE[2]\textbf{case} negative:~$\mathbf{v} \gets \left( \mathbf{1}_m || \mathbf{1}_{\bar{m}}\right), m\neq\bar{m}$
\INDSTATE[2]\textbf{case} neutral:~$\mathbf{v} \gets \left( \mathbf{1}_m || \mathbf{0}\right)$
\INDSTATE[1]$(\mathbf{a}_0,\dots,\mathbf{a}_{L-1}), (\mathcal{L}_0,\dots,\mathcal{L}_{L-1}) \gets \textsc{FF-Net}\text{($\boldsymbol{\theta}^{\mathrm{e}}$, $\mathbf{v}$, t)}$
\INDSTATE[1]$\mathbf{x} \gets \textsc{Normalize}\text{($\mathbf{a}_{L-1}$)} \text{ or }\textsc{Quantize}\text{($\mathbf{a}_{L-1}$)}$
\INDSTATE[1]$\mathbf{y} \gets \textsc{Channel}\text{($\mathbf{x}$)}$
\INDSTATE[1]$(\mathbf{a}_{L},\dots,\mathbf{a}_{L+K-1}), (\mathcal{L}_L,\dots,\mathcal{L}_{L+K-1}) \gets \textsc{FF-Net}\text{($\boldsymbol{\theta}^\mathrm{d}$, $\mathbf{y}$, t)}$
\INDSTATE[1]$\mathbf{p} \gets \mathrm{Softmax}\left( c_{\mathbf{\boldsymbol{\kappa}}} \left(\mathbf{a}_{L},\dots,\mathbf{a}_{L+K-1}\right) \right)$
\INDSTATE[1]$\mathcal{L}_\mathrm{CCE} \gets \mathrm{CCE}\left(\mathbf{p}, \mathbf{1}_m\right)$
\INDSTATE[1]\textbf{return}~$(\mathcal{L}_0,\dots,\mathcal{L}_{L+K-1}, \mathcal{L_\mathrm{CCE}})$
\INDSTATE[0]\textbf{end function}
\INDSTATE[0]
\INDSTATE[0]\textbf{function}~\textsc{FF-Net}($\boldsymbol\theta$, $\mathbf{x}_0$, $t$):
\INDSTATE[1]$\mathbf{x_0} \gets  \mathbf{x_0}/\lVert\mathbf{x}_0 \rVert_2$
\INDSTATE[1]\textbf{for} each layer~$i$ in $L$-layer network:
\INDSTATE[2]$\mathbf{a}_i \gets f_{\mathbf{\theta}_i}(\mathbf{x}_i)$
\INDSTATE[2]$g_i \gets \lVert\mathbf{a}_i \rVert_2^2$
\INDSTATE[2]\textbf{switch} $t$ \textbf{do}
\INDSTATE[3]\textbf{case} positive:~$\mathcal{L}_i \gets \zeta\left(- \left(g_i - \tau_i\right)\right)$
\INDSTATE[3]\textbf{case} negative:~$\mathcal{L}_i \gets \zeta\left(\left(g_i - \tau_i\right)\right)$
\INDSTATE[3]\textbf{case} neutral:~$\mathcal{L}_i \gets \emptyset$
\INDSTATE[2]$\mathbf{a}_i \gets \mathbf{a}_i/\lVert\mathbf{a}_i \rVert_2$
\INDSTATE[2]$\mathbf{x}_{i+1} \gets \mathbf{a}_i$
\INDSTATE[1]\textbf{end for}
\INDSTATE[1]\textbf{return}~$(\mathbf{a}_0,\dots ,\mathbf{a}_{L-1}), (\mathcal{L}_0,\dots,\mathcal{L}_{L-1})$
\INDSTATE[0]\textbf{end function}
\end{algorithmic}
\label{alg:ff_network_outer}
\end{algorithm}
\vspace{-1.0cm}
\end{figure}

In the following, we will briefly describe the training procedure, which is based on the initial proposal from~\cite{hinton2022}. Algorithms~\ref{alg:training} and~\ref{alg:autoencoder} outline the training in more detail. Assume an encoder and a decoder network with~$L$ and~$K$ fully connected layers, respectively. The performance of each layer is quantified by a goodness measure~$g_i = \lVert\mathbf{a}_i \rVert_2^2$. The optimization of the parameters~$\boldsymbol{\theta}_i$ is achieved via stochastic gradient descent (SGD), employing the learning rate~$\gamma_f$, weight decay~$\lambda_f$, and momentum~$\mu_f$. The corresponding loss function depends on whether a positive or negative sample is processed. It is defined as
\begin{equation}
    \mathcal{L}_i(g_i,\tau_i) = 
    \begin{cases}
    \zeta\left(- \left(g_i - \tau_i\right)\right)& \text{if positive sample,}\\
    \zeta\left(g_i - \tau_i\right)& \text{if negative sample}
    \end{cases}
\end{equation}
where~$\zeta(x)= \log(1+e^x)$ denotes the softplus function and~$\tau_i$ is a threshold value that we statically assign with the output width of the current layer. Intuitively, this loss aims to increase the network activities, quantified by the goodness metric, (above~$\tau_i$) for positive data and decrease it (below~$\tau_i$) for negative data. At the end of each layer, the outputs are normalized with respect to the~$l^2$ norm in order to process only the relative activities from one neuron to the next.

Alongside the encoder and decoder layers that are trained via the FF algorithm, the decoder additionally employs a single classification layer~$c_{\boldsymbol{\kappa}}(\cdot)$ with a softmax output that learns how to associate the decoder's network activities~$\mathbf{a}_i$ with the originally encoded message~$m$, where~$L \leq i < L + K -1$. During this step, the rest of the network is provided with neutral input samples to generate the decoder activities. The classifier is trained via SGD with the hyperparameters~$\gamma_c$,~$\lambda_c$, and~$\mu_c$ using the CCE loss. This also requires the availability of the correct labels at the output of the classifier, which could practically be accomplished by a set of pilot messages for which the ground truth is known. Note that the classifier tuning does not break the requirement of single-layer optimization.

\section{Numerical Results}
This work aims to provide an initial performance assessment of the proposed coding scheme, which is typically evaluated with respect to the average probability of decoding error~$P_e = \mathrm{Pr(\hat{m}\neq m)}$, where~$\hat{m}$ denotes the decoding result, i.e., the index of the largest element in~$\mathbf{p}$. This error probability is empirically estimated by the BLER via Monte Carlo simulations. We consider a real-valued Rayleigh block fading (RBF) channel~$\left(\mathcal{X}, p_{Y|X},\mathcal{Y}\right)$ given for a sequence of~$n$ consecutive symbols by
\begin{equation}
    Y_i = HX_i + N_i 
\end{equation}
where~$N_i$ is a zero-mean Gaussian random variable with variance~$\sigma^2 = \left( 2RE_b/N_0 \right)^{-1}$ and~$E_b/N_0$ is the per-bit energy to noise power spectral density, which we also refer to as the signal-to-noise ratio (SNR). The random variable~$H$ follows a Rayleigh distribution that models the magnitude of the channel's fading coefficients. The SNR is kept constant at~$E_b/N_0=\SI{5}{\dB}$ during training as previous studies on BP-based autoencoders~\cite{oshea2017} show an adequate ability of generalization to other SNR domains. Moreover, we define the code rate as~$R=k/n$, which we will fix at~$R=4/7$ throughout all experiments. Note that these short-length codes could potentially be extended to higher blocklengths by using concatenated code construction schemes as proposed in~\cite{gunluConcatenatedClassicNeural2023a}. Further training hyperparameters are disclosed in Appendix~\ref{appendix:network_config_hyperparameters}.

\vspace{-0.15cm}
\subsection{Joint Coding and Modulation}
\label{subsec:joint_coding_modulation}

We will first consider an autoencoder whose encoder comprises both coding and modulation stages, having no restrictions on the domain of its output in~$\mathbb{R}^n$ except for an average power normalization, i.e., $\mathbb{E}(|x_i|^2)\leq 1$. In addition to the seminal autoencoder from~\cite{oshea2017}, we incorporate results from deploying the RL-enabled system with a noiseless feedback link~\cite{aoudia2019}, where encoder and decoder are trained in an alternating manner to circumvent BP through the channel. In this model-free algorithm, the decoder first performs~$10$ optimization rounds using the true gradient, after which the parameters of the encoder are tuned in~$10$ optimization rounds using an approximation of the gradient. This approximation is enabled by a distortion of the encoder output with additive Gaussian noise following~$\mathcal{N}(0,\sigma_\mathrm{RL})$, where we selected~$\sigma_\mathrm{RL}=0.1$ to control the amount of exploration within the stochastic policy of the RL algorithm.

\begin{table}[t]
\vspace{0.2cm}
\caption{BLER for FF networks of varying size: The networks consist of~$L$ encoder and~$K$ decoder layers, each of width~$W$ (excluding the classifier). The BLER is measured at~$E_b/N_0=\SI{7}{\dB}$.}
    \begin{minipage}{.48\linewidth}
        \centering
        \subcaption{$W=16$}
        \label{tab:ff_bler_width_16}
        \begin{tabular}{ c c H H H }
            \toprule
            \multicolumn{2}{c}{BLER} & \multicolumn{3}{c}{$L$} \\
            \multicolumn{2}{c}{in $10^{-3}$} & \multicolumn{1}{c}{2} & \multicolumn{1}{c}{3} & \multicolumn{1}{c}{4} \\
            \midrule
            & 2 & 9.1 & 5.6 & 37.0 \\
            $K$ & 3 & 9.3 & 5.4 & 36.8 \\
            & 4 & 8.9 & 5.4 & 35.8 \\
            \bottomrule
        \end{tabular}
    \end{minipage}
    \hfill
    \begin{minipage}{.48\linewidth}
        \centering
        \subcaption{$W=80$}
        \label{tab:ff_bler_width_80}
        \begin{tabular}{ c c H H H }
            \toprule
            \multicolumn{2}{c}{BLER} & \multicolumn{3}{c}{$L$} \\
            \multicolumn{2}{c}{in $10^{-3}$} & \multicolumn{1}{c}{2} & \multicolumn{1}{c}{3} & \multicolumn{1}{c}{4} \\
            \midrule
            & 2 & 2.6 & 4.9 & 1.5 \\
            $K$ & 3 & 2.4 & 5.0 & 1.5 \\
            & 4 & 2.6 & 5.3 & 1.3 \\
            \bottomrule
        \end{tabular}
    \end{minipage}
    \label{tab:ff_bler_vs_network_size}
    \vspace{-0.55cm}
\end{table}

Before comparing the different training algorithms, it needs to be studied what impact the network capacity in terms of depth and layer width will have on the overall performance. In Tables~\ref{tab:ff_bler_width_16} and~\ref{tab:ff_bler_width_80}, the BLER is depicted for varying numbers of encoder layers~$L$, decoder layers~$K$, and neurons per layer~$W$. Note that the input size of the single-layer classifier also depends on the number and width of the previous decoder layers. In order to obtain more nuanced results, these evaluations were performed at a slightly higher SNR than the one applied during training. It can be observed that FF networks generally require wider layers to achieve an adequate performance. This stands in contrast to BP-based autoencoders, where a similarly significant improvement could not be observed for deeper and wider networks. Moreover, the overall performance of the FF autoencoder seems to be more dependent on the complexity of the encoder as the BLER range only slightly improves for increasing~$K$. If the encoder is structurally incapable of learning a robust representation of the input message, a decoder of any size will not be able to correctly decode the received block. As our focus is to examine the potential capabilities of the proposed FF autoencoder architecture, throughout the rest of this work, all experiments will be conducted using the most complex implementation with~$L=K=4$ and~$W=80$ as it provides the best performance.

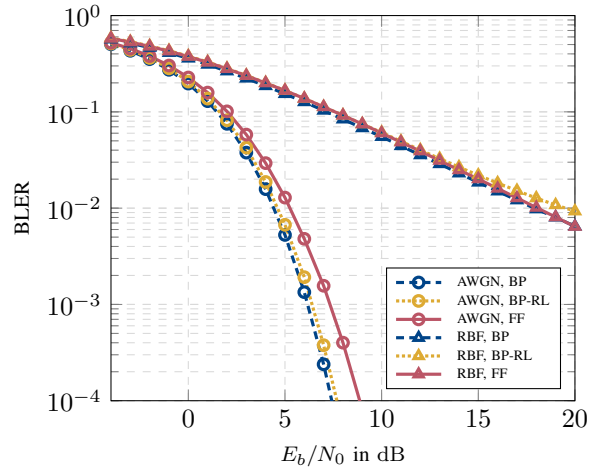
\begin{figure}[t!]
    \centering
    \begin{tikzpicture}[trim axis left, trim axis right]
    \begin{semilogyaxis}[
    		legend style={nodes={scale=0.53, transform shape}, legend columns=1, column sep=0.15cm},
            legend cell align={left},
            legend pos=south east,
    		xmin=-4, xmax=20,
    		ymin=1e-4, ymax=1,
    		grid=both,
    		xlabel={\footnotesize{$E_b/N_0$} in~$\si{dB}$},
    		ylabel={\footnotesize{BLER}},
    		grid style={dashed,gray!30},
            scale=0.8
    	]
     
        \addplot[color=plot0, mark=o, densely dashed, mark options={solid}, very thick] table [col sep=comma, x=EbN0_dB, y=bler]
            {plots/data_new/bp_awgn/bler_bp.csv};
        \addplot[color=plot1, mark=o, densely dotted, mark options={solid}, very thick] table [col sep=comma, x=EbN0_dB, y=bler]
            {plots/data_new/bp_rl_awgn/bler_bp.csv};
        \addplot[color=plot2, mark=o, solid, mark options={solid}, very thick] table [col sep=comma, x=EbN0_dB, y=bler]
            {plots/data_new/ff_awgn/bler_ff.csv};
        \addplot[color=plot0, mark=triangle, densely dashed, mark options={solid}, very thick] table [col sep=comma, x=EbN0_dB, y=bler] 
            {plots/data_new/bp_fading/bler_bp.csv};
        \addplot[color=plot1, mark=triangle, densely dotted, mark options={solid}, very thick] table [col sep=comma, x=EbN0_dB, y=bler] 
            {plots/data_new/bp_rl_fading/bler_bp.csv};
        \addplot[color=plot2, mark=triangle, solid, mark options={solid}, very thick] table [col sep=comma, x=EbN0_dB, y=bler] 
            {plots/data_new/ff_fading/bler_ff.csv};
    	
    	\legend{
            {\small{AWGN, BP}},
            {\small{AWGN, BP-RL}},
            {\small{AWGN, FF}},
            {\small{RBF, BP}},
            {\small{RBF, BP-RL}},
            {\small{RBF, FF}},
    	}
    \end{semilogyaxis}
    \end{tikzpicture}
    \caption{BLER over~$E_b/N_0$ of the continuous-output autoencoders for the AWGN and RBF channels.}
    \label{fig:bler_awgn_rbf}
    \vspace{-0.5cm}
\end{figure}

In Fig.~\ref{fig:bler_awgn_rbf}, we compare the BLER over~$E_b/N_0$ for the autoencoder trained with the BP, BP-RL, and FF algorithms for the aforementioned channel models. In the simple additive white Gaussian noise (AWGN) scenario, i.e.,~$H=1$, the BLER of the FF autoencoder ranges close to those of the BP- and BP-RL-based systems; however, its performance deteriorates when increasing~$E_b/N_0$, resulting in an SNR gap of around~$\SI{1}{\dB}$. In the case of RBF, which typically results in less steep BLER curves due to more perturbations introduced by the channel, the FF autoencoder is able to compete with the other approaches. This hints at potential limitations of the FF algorithm in learning more nuanced representations of the input bits that predominantly become more crucial in lower BLER regimes.

\vspace{-.2cm}
\subsection{Quantized Encoder Outputs}
\begin{figure}[t!]
    \centering
    \begin{tikzpicture}[trim axis left, trim axis right]
    \begin{semilogyaxis}[
    		legend style={nodes={scale=0.53, transform shape}, legend columns=1, column sep=0.15cm},
            legend cell align={left},
            legend pos=south west,
    		xmin=-4, xmax=20,
    		ymin=1e-4, ymax=1,
    		grid=both,
    		xlabel={\footnotesize{$E_b/N_0$} in~$\si{dB}$},
    		ylabel={\footnotesize{BLER}},
    		grid style={dashed,gray!30},
            scale=0.8
    	]
     
        \addplot[color=plot0, mark=o, densely dashed, mark options={solid}, very thick] table [col sep=comma, x=EbN0_dB, y=bler]
            {plots/data_new/bp_awgn_hard_decision/bler_bp.csv};
        \addplot[color=plot1, mark=o, densely dotted, mark options={solid}, very thick] table [col sep=comma, x=EbN0_dB, y=bler]
            {plots/data_new/bp_rl_awgn_hard_decision/bler_bp.csv};
        \addplot[color=plot2, mark=o, solid, mark options={solid}, very thick] table [col sep=comma, x=EbN0_dB, y=bler]
            {plots/data_new/ff_awgn_hard_decision/bler_ff.csv};
        \addplot[color=plot0, mark=triangle, densely dashed, mark options={solid}, very thick] table [col sep=comma, x=EbN0_dB, y=bler] 
            {plots/data_new/bp_fading_hard_decision/bler_bp.csv};
        \addplot[color=plot1, mark=triangle, densely dotted, mark options={solid}, very thick] table [col sep=comma, x=EbN0_dB, y=bler] 
            {plots/data_new/bp_rl_fading_hard_decision/bler_bp.csv};
        \addplot[color=plot2, mark=triangle, solid, mark options={solid}, very thick] table [col sep=comma, x=EbN0_dB, y=bler] 
            {plots/data_new/ff_fading_hard_decision/bler_ff.csv};
    	
    	\legend{
            {\small{AWGN, BP}},
            {\small{AWGN, BP-RL}},
            {\small{AWGN, FF}},
            {\small{RBF, BP}},
            {\small{RBF, BP-RL}},
            {\small{RBF, FF}},
    	}
    \end{semilogyaxis}
    \end{tikzpicture}
    \caption{BLER over~$E_b/N_0$ of the autoencoders with quantized-output encoder for the AWGN and RBF channels.}
    \label{fig:bler_awgn_rbf_quantized}
    \vspace{-0.5cm}
\end{figure}
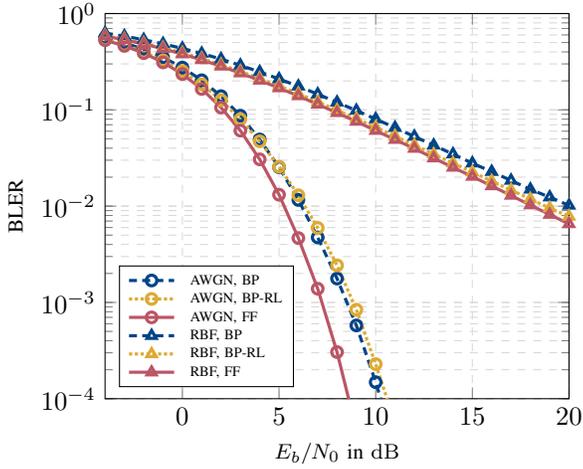

In certain communications systems, the encoder can be required to map its output to a fixed set of modulation symbols, guaranteeing a hard power constraint such as Binary Phase Shift Keying (BPSK) modulation. Another useful property of the separation of coding and modulation stages is the ability to perform a more precise characterization of coding gains achieved by the learned autoencoder. However, in the simplest example of a BPSK-enforced autoencoder, quantization of the encoder's output via~$x_j = \sign(a_j)$, where~$j\in\left[0,N-1\right]$, is not differentiable in terms of having a zero gradient almost everywhere, which poses an obstacle for the application of the BP algorithm. To overcome this issue, \cite{jiangTurboAutoencoderDeep2019a} proposed to use a surrogate model that implements the saturated straight through estimator (STE) \cite{bengioEstimatingPropagatingGradients2013}, which propagates the gradient in the backward path as
\vspace{-0.2cm}
\begin{align}
    \frac{\partial x_j}{\partial a_j} = 
    \begin{cases}
    1& \text{if } \lvert a_j\rvert < 1\text{,}\\
    0& \text{otherwise.}
    \end{cases}
\end{align}
In the following, we repeat the numerical characterization of the systems with respect to the BLER, depicted in Fig.~\ref{fig:bler_awgn_rbf_quantized}. The BP autoencoder implements the STE-based backward path, while the BP-RL system incorporates the non-differentiable quantization operation as part of the channel. For the FF autoencoder, no adjustments need to be made as it does not require a fully differentiable path through the system. For the AWGN channel, it can be observed that while the FF autoencoder retains its performance, both the BP- and BP-RL autoencoders clearly increase their respective BLER in comparison with the non-quantized system, which is consistent with the findings from~\cite{jiangTurboAutoencoderDeep2019a}. The FF autoencoder proves to be superior to the BP-based approach in the RBF scenario as well, while the BP-RL autoencoder is able to almost close the gap. This shows that the STE-enabled surrogate model used by the BP autoencoder insufficiently approximates the backward gradient flow through the quantizer.

\subsection{Convergence Rate}
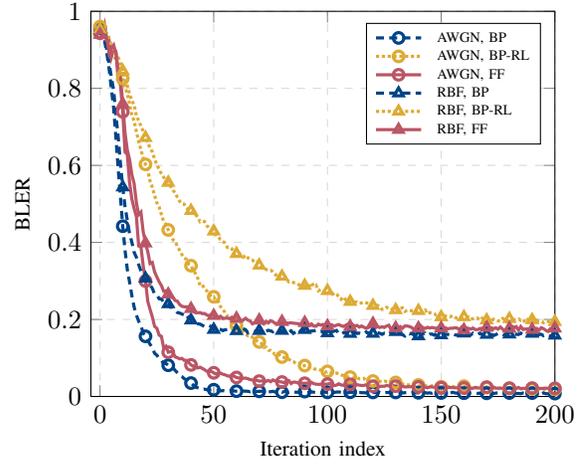
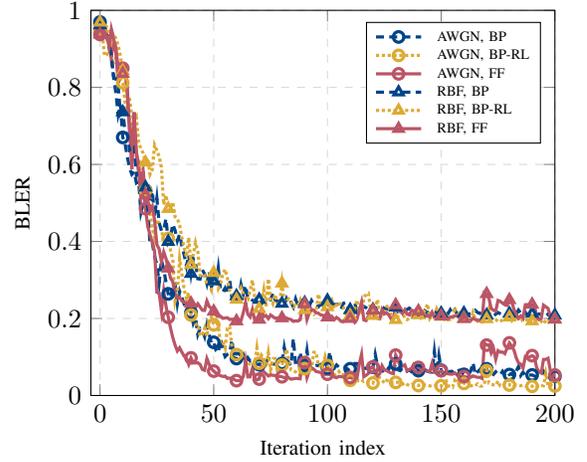
\begin{figure}[t!]
\centering
    \begin{subfigure}[t]{0.5\textwidth}
    \centering
        \begin{tikzpicture}[trim axis left, trim axis right]
            \begin{axis}[
                    legend style={nodes={scale=0.53, transform shape}, legend columns=1, column sep=0.15cm},
                    legend cell align={left},
                    legend pos=north east,
                    xmin=-4, xmax=200,
                    ymin=0, ymax=1.0,
                    grid=both,
                    xlabel={\footnotesize{Iteration index}},
                    ylabel={\footnotesize{BLER}},
                    grid style={dashed,gray!30},
                    scale=0.8
                ]
             
                \addplot[color=plot0, mark=o, mark repeat={10}, densely dashed, mark options={solid}, very thick] table [col sep=comma, x=iter, y=bler]
                    {plots/data_new/bp_awgn/bler_per_iter_filtered_10.csv};
                \addplot[color=plot1, mark=o, mark repeat={10}, densely dotted, mark options={solid}, very thick] table [col sep=comma, x=iter, y=bler]
                    {plots/data_new/bp_rl_awgn/bler_per_iter.csv};
                \addplot[color=plot2, mark=o, mark repeat={10}, solid, mark options={solid}, very thick] table [col sep=comma, x=iter, y=bler]
                    {plots/data_new/ff_awgn/bler_per_iter_filtered_5.csv};
                \addplot[color=plot0, mark=triangle, mark repeat={10}, densely dashed, mark options={solid}, very thick] table [col sep=comma, x=iter, y=bler]
                    {plots/data_new/bp_fading/bler_per_iter_filtered_10.csv};
                \addplot[color=plot1, mark=triangle, mark repeat={10}, densely dotted, mark options={solid}, very thick] table [col sep=comma, x=iter, y=bler]
                    {plots/data_new/bp_rl_fading/bler_per_iter.csv};
                \addplot[color=plot2, mark=triangle, mark repeat={10}, solid, mark options={solid}, very thick] table [col sep=comma, x=iter, y=bler]
                    {plots/data_new/ff_fading/bler_per_iter_filtered_5.csv};
                
                \legend{
                    {\small{AWGN, BP}},
                    {\small{AWGN, BP-RL}},
                    {\small{AWGN, FF}},
                    {\small{RBF, BP}},
                    {\small{RBF, BP-RL}},
                    {\small{RBF, FF}},
                }
            \end{axis}
            \end{tikzpicture}
        \caption{Continuous-output encoder.}
        \label{fig:convergence_continuous}
    \end{subfigure}
    ~
    \begin{subfigure}[t]{.5\textwidth}
        \centering
        \begin{tikzpicture}[trim axis left, trim axis right]
        \begin{axis}[
        		legend style={nodes={scale=0.53, transform shape}, legend columns=1, column sep=0.15cm},
                legend cell align={left},
                legend pos=north east,
        		xmin=-4, xmax=200,
        		ymin=0, ymax=1.0,
        		grid=both,
        		xlabel={\footnotesize{Iteration index}},
        		ylabel={\footnotesize{BLER}},
        		grid style={dashed,gray!30},
                scale=0.8
        	]
         
            \addplot[color=plot0, mark=o, mark repeat={10}, densely dashed, mark options={solid}, very thick] table [col sep=comma, x=iter, y=bler]
                {plots/data_new/bp_awgn_hard_decision/bler_per_iter_filtered_10.csv};
            \addplot[color=plot1, mark=o, mark repeat={10}, densely dotted, mark options={solid}, very thick] table [col sep=comma, x=iter, y=bler]
                {plots/data_new/bp_rl_awgn_hard_decision/bler_per_iter.csv};
            \addplot[color=plot2, mark=o, mark repeat={10}, solid, mark options={solid}, very thick] table [col sep=comma, x=iter, y=bler]
                {plots/data_new/ff_awgn_hard_decision/bler_per_iter_filtered_5.csv};
            \addplot[color=plot0, mark=triangle, mark repeat={10}, densely dashed, mark options={solid}, very thick] table [col sep=comma, x=iter, y=bler]
                {plots/data_new/bp_fading_hard_decision/bler_per_iter_filtered_10.csv};
            \addplot[color=plot1, mark=triangle, mark repeat={10}, densely dotted, mark options={solid}, very thick] table [col sep=comma, x=iter, y=bler]
                {plots/data_new/bp_rl_fading_hard_decision/bler_per_iter.csv};
            \addplot[color=plot2, mark=triangle, mark repeat={10}, solid, mark options={solid}, very thick] table [col sep=comma, x=iter, y=bler]
                {plots/data_new/ff_fading_hard_decision/bler_per_iter_filtered_5.csv};
        	
        	\legend{
                {\small{AWGN, BP}},
                {\small{AWGN, BP-RL}},
                {\small{AWGN, FF}},
                {\small{RBF, BP}},
                {\small{RBF, BP-RL}},
                {\small{RBF, FF}},
        	}
        \end{axis}
        \end{tikzpicture}
        \caption{Quantized-output encoder.}
        \label{fig:convergence_quantized}
    \end{subfigure}
    \caption{BLER over training iterations of the continuous-output (a) and the quantized-output (b) encoder for~$E_b/N_0=\SI{5}{\dB}$.}
    \vspace{-0.5cm}
\end{figure}

The sample complexity of a training algorithm describes the number of samples required to achieve convergence with respect to the target loss or error function. Figs.~\ref{fig:convergence_continuous} and~\ref{fig:convergence_quantized} illustrate the evolution of the BLER over the training iterations at a fixed~$E_b/N_0=\SI{5}{\dB}$ for the models trained with the BP, BP-RL, and FF algorithms for both continuous-encoder and quantized-encoder systems, respectively. For the RL-based approach, one iteration accounts for~$10$ alternating optimization rounds of transmitter and receiver training. Similarly, the FF algorithm performs one positive and one negative pass during each iteration. Therefore, in order to ensure a fair comparison, we evaluate every~$10$th iteration for the BP algorithm and every~$5$th iteration for the FF algorithm.

In the continuous-encoder scenario, more iterations are required for RL to reduce the BLER to the same range as classical BP for the AWGN and RBF channels, as it has already been observed in~\cite{aoudia2019}. The FF algorithm clearly outperforms RL as it converges as fast as BP. In contrast, the convergence curves for the systems involving the quantized encoder are less stable than for the continuous case for all training algorithms. This shows that the non-differentiable operation poses a challenge for all training approaches equally. However, the FF-based autoencoder is able to reach the target loss more quickly, while the BP and BP-RL-trained systems display a similar convergence behavior.

While the learning rates used during training (except for the FF classifier) are the same, we acknowledge that the increased network capacity of the FF autoencoder may have a major impact on its convergence speed. However, as pointed out in Subsection~\ref{subsec:joint_coding_modulation}, FF networks generally require a larger parameter space to achieve comparable performance to their BP-based counterparts. Moreover, during our experiments, we found that a comparably large network for the BP and BP-RL autoencoders does not lead to significant decreases in loss, and for the quantized-output encoder, it even leads to an increase in loss.

\subsection{Hardware Complexity Discussions}
Although this work is mainly simulation-based, from an algorithmic perspective, we can derive certain implications that the FF-based autoencoder architectures will have on potential hardware implementations. As mentioned, the proposed FF-trained networks require a larger network capacity to compete with BP performance. This, unfortunately, leads to an increase in the overall computational overhead. However, the true gains of the FF approach will lie in breaking the backward lock, i.e., reducing the processing time introduced in the backward path, when layers would wait for the gradients from succeeding layers to arrive, and in eliminating the need to store all backward derivatives. 

To quantify the impact on processing time, consider a neural network of $N$ layers, where each forward and backward pass per layer would consume an equal amount of time. In this case, the update of all network parameters would require~$2N$ time units using the BP algorithm, while FF learning could achieve the same within only~$N+1$ steps.

Moreover, we highlight the memory savings, taking the BP network used throughout this work as an example. For this configuration, the complete autoencoder would have to allocate memory for the gradients with respect to~$791$ parameters, already considering per-node and per-batch accumulation, i.e., not tracking the derivative of every function along the way and for every data sample within the batch. In contrast, the FF algorithm does not require these gradients to be stored, as they are computed and directly consumed to adjust the parameters in every single layer. Especially in digital hardware, the data traffic due to memory operations forms a bottleneck for both the computation speed and the energy consumption~\cite{nguyenApproximateMemoryArchitecture2020}. Therefore, the algorithmic properties of FF learning could alleviate these constraints and enable efficient implementations of deep learning even on very low-power edge devices.

\section{Conclusion}
In this work, we designed an end-to-end learned autoencoder for wireless communications whose training is enabled by the FF algorithm. We illustrated that this design is able to compete with existing models based on BP for both the AWGN and RBF channels and even outperform them in a scenario with an enforced, non-differentiable quantization stage. Moreover, we showed that the considered FF networks converge with comparable speed or even faster than similarly performing networks trained with BP. Although the FF algorithm exhibits some deficiencies, such as an increased sensitivity to the training hyperparameters, it is a suitable candidate to overcome the energy efficiency and memory consumption problems of neural codes. Thus, it enables a more efficient hardware implementation of neural networks and applications to non-differentiable channels without the need for a feedback link.

Future studies will consider applications of the proposed FF autoencoders to more complex, non-differentiable channel models. As the research on FF learning is fairly immature, potential extensions of the algorithm towards more sophisticated loss function design and layer collaboration techniques could further improve its performance.

\bibliographystyle{IEEEtran}
\bibliography{IEEEabrv,refs}

\appendices
\section{Network Configuration and Hyperparameters}\label{appendix:network_config_hyperparameters}
As the training of the FF autoencoder is a very delicate process, we provide the hyperparameters used in this work's experiments in Table~\ref{tab:hyperparameters} for the sake of reproducibility. All the weights of the BP-trained networks and the classification layer of the FF autoencoder were initialized using a Kaiming uniform distribution~\cite{heDelvingDeepRectifiers2015a}; the biases were set to zero. Note that an extensive hyperparameter search has yet to be conducted.


\begin{table}[h!]
\caption{Hyperparameters for the autoencoder training using the BP, BP-RL and FF algorithms.}
\label{tab:hyperparameters}
\footnotesize
\centering
\begin{tabular}{l c c c}
\toprule
Parameter&BP&BP-RL&FF \\
\midrule
Number of encoder layer~$L$&~$2$&~$2$&~$4$\\
Number of decoder layer~$K$&~$2$&~$2$&~$4$\\
Width of hidden layers~$W$&~$16$&~$16$&~$80$\\
Max. number of iterations &~$5000$&~$18000$&~$8200$\\
Batch size &~$250$&~$250$&~$250$\\
Optimizer & Adam&Adam&SGD\\
Learning rate~$\gamma$ &~$0.001$&~$0.001$&---\\
Learning rate (FF network)~$\gamma_f$ &---&---&~$0.001$\\
Learning rate (classifier)~$\gamma_c$ &---&---&~$0.005$\\
Weight decay (FF network)~$\lambda_f$ &---&---&~$0.0003$\\
Weight decay (classifier)~$\lambda_c$ &---&---&~$0.003$\\
Momentum (FF network)~$\mu_f$ &---&---&~$0.9$\\
Momentum (classifier)~$\mu_c$ &---&---&~$0.9$\\
\bottomrule
\end{tabular}
\end{table}

\end{document}